\documentstyle[epsf,aps]{revtex}
\renewcommand{\thefootnote}{\fnsymbol{footnote}}
\begin{document}
\newcommand{\be}{\begin{eqnarray}}
\newcommand{\dlq}{\lq\lq}
\newcommand{\ee}{\end{eqnarray}}
\newcommand{\ben}{\begin{eqnarray*}}
\newcommand{\een}{\end{eqnarray*}}
\newcommand{\stackeven}[2]{{{}_{\displaystyle{#1}}\atop\displaystyle{#2}}}
\newcommand{\lsim}{\stackeven{<}{\sim}}
\newcommand{\gsim}{\stackeven{>}{\sim}}
\renewcommand{\baselinestretch}{1.0}
\newcommand{\as}{\alpha_s}
\def\eq#1{{Eq.~(\ref{#1})}}
\def\fig#1{{Fig.~\ref{#1}}}
\begin{flushright}
NT@UW-00-031
\end{flushright}
\vspace*{1cm} 
\setcounter{footnote}{1}
\begin{center}
{\Large\bf Classical Initial Conditions for Ultrarelativistic
Heavy Ion \\ ~~ \\ Collisions}
\\[1cm]
Yuri V.\ Kovchegov \\ ~~ \\
{\it Department of Physics, University of Washington, Box 351560} \\ 
{\it Seattle, WA 98195-1560} \\ ~~ \\ ~~ \\
\end{center}
\begin{abstract} 
We construct an analytical expression for the distribution of gluons
in the state immediately following a heavy ion collision in the
quasi--classical limit of QCD given by McLerran--Venugopalan
model. The resulting gluon number distribution function includes the
effects of all multiple rescatterings of gluons with the nucleons of
both colliding nuclei. The typical transverse momentum $k_\perp$ of
the produced gluons is shown to be of the order of the saturation
scale of the nuclei $Q_s$, as predicted by Mueller. We analyze the
properties of the obtained distribution and demonstrate that due to
multiple rescatterings it remains finite (up to logarithms of
$k_\perp$) in the soft transverse momentum limit of $k_\perp \,
\ll \, Q_s$ unlike the usual perturbative initial conditions given by
collinear factorization. We calculate the total number of produced
gluons and show that it is proportional to the total number of gluons
inside the nuclear wave function before the collision with the
proportionality coefficient $c \, \approx \, 2 \, \ln 2$.
\end{abstract}
\renewcommand{\thefootnote}{\arabic{footnote}}
\setcounter{footnote}{0}

\section{Introduction}

Relativistic Heavy Ion Collider (RHIC) at Brookhaven National
Laboratory has recently became operational with new data already being
produced \cite{pho,star}. One of the challenges facing theoretical
heavy ion community is the correct interpretation of the newly
obtained data. It has been conjectured that in a high energy heavy ion
collision a thermalized state of quarks and gluons, usually referred
to as quark--gluon plasma (QGP), may be produced \cite{qgp}, allowing
us to explore the properties of QCD matter under extreme
conditions. Understanding the experimental signatures of QGP is a very
difficult task.  The first important step in that direction is finding
a correct description of the state of gluons and quarks immediately
following a heavy ion collision but preceding the anticipated onset of
thermalization. Distribution of partons in this state has become known
as the initial conditions for QGP formation. The succeeding evolution
of this quark-gluon system and its possible thermalization can be
studied by Monte-Carlo simulations \cite{gmgw}, by invoking the
formalism of transport theory \cite{dg} or by analytical estimates of
\cite{bmss,Mueller2}.

One may try to model the initial condition by considering pairwise
interactions of nucleons in the colliding nuclei. The particle
production would be described by applying collinear factorization
formalism to each individual nucleon-nucleon collision and
superimposing the results \cite{coll}. Unfortunately the approach has
several shortcomings. In the integration over the transverse momentum
of one of the produced partons one is forced to introduce an infrared
cutoff $p_0$ in order to make the integral finite. The resulting
production cross section depends very strongly on the cutoff $p_0$
which makes it difficult to determine the initial conditions with high
precision. Another (related) problem of the purely collinear
factorization approach is that it does not include the effects of
multiple rescatterings of the produced partons with each other and
other nucleons, which seem to be, at least intuitively, characteristic
to heavy ion collisions where the number of scattering particles is
large. This phenomenon is intimately related to the problems of
nuclear shadowing and higher twist resummation. There have been made
several attempts to correct the problem by redefining the nucleons'
parton distributions to include the nuclear saturation effects and
then using collinear factorization mechanism to describe particle
production \cite{coll2}.

An interesting model of nuclear collisions has been proposed recently
by McLerran and Venugopalan \cite{mv}, which was designed to both
explain the nuclear shadowing phenomenon and to provide us with the
strategy of deriving the initial state parton distributions free of
problems typical to collinear factorization approaches. The model is
based on the observation that at very high energies the parton
densities in large nuclei reach saturation \cite{GLR} and the number
of partons becomes very large. Then due to a large number of color
charge sources the gluon emission can be described by the solution of
the classical Yang-Mills equations of motion \cite{mv}.

\begin{figure}
\begin{center}
\epsfxsize=8cm
\leavevmode
\hbox{ \epsffile{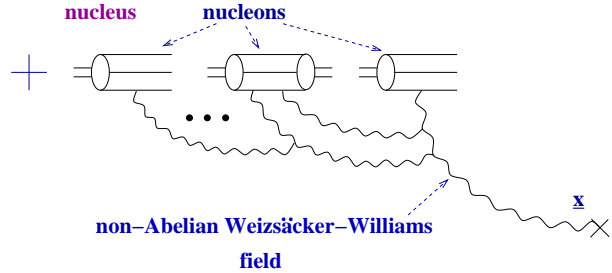}}
\end{center}
\caption{Non-Abelian Weizs\"{a}cker-Williams field of a large nucleus, 
as derived in \protect\cite{me,jklw,me2}. }
\label{wwfig}
\end{figure}

To describe the saturation philosophy let us start by considering a
small-$x$ gluon distribution function of a single ultrarelativistic
nucleus. Let us suppose that Bjorken $x$ is small, $x \, \ll \, 1$,
but not too small for quantum corrections to become important
\cite{M3,bdmps}. The quantum corrections at small $x$ bring in the
powers of $\as \ln 1/x$ and are resummed for the case of one
rescattering by the BFKL equation \cite{BFKL}. If $\as \ln 1/x \, \ll
\, 1$ we can describe the gluon distribution in the nuclear wave function 
including multiple rescatterings on the nucleons by the classical
gluon field calculated in light cone gauge \cite{mv}. The field has
been found in \cite{me,jklw} and was dubbed the non-Abelian
Weizs\"{a}cker-Williams field of a large nucleus. In the $A_+ = 0$
light cone gauge (with the nucleus moving in the ``plus'' direction
along the light cone) the field has only transverse components which
are given by \cite{me,jklw}
\be\label{ww}
{\underline A}^{WW} ({\underline x},x_-) \ = \ \int \ d^2 z \ d z_-
\theta (z_- - x_-) \ {\hat \rho}^a ({\underline z},z_-) \
\frac{{\underline x} - {\underline z}}{|{\underline x} - {\underline z}|^2} 
\ S_0 ({\underline x},z_-) T^a S_0^{-1} ({\underline x},z_-)
\ee
with the gauge rotation matrix given by a path-ordered integral
\be
S_0 ({\underline x},x_-) \ = \ \mbox{P} \exp \left( i g T^a \int \ d^2
z \ d z_- \theta (z_- - x_-) \ {\hat \rho}^a ({\underline z},z_-) \ \ln
(|{\underline x} - {\underline z}| \mu) \right).
\ee
${\hat \rho}^a$ is a color charge density operator normalized
according to
\be\label{dcorr}
\left< {\hat \rho}^a ({\underline x},x_-) \ {\hat \rho}^b ({\underline y},y_-) 
\right> \ = \ \frac{\as}{2 N_c \pi} \, \rho ({\underline x},x_-) \, 
\delta (x_- - y_-) \,  \delta^2 ({\underline x} - {\underline y}) \, \delta^{ab}
\ee
with $\rho ({\underline x},x_-)$ the normal nuclear density in the
infinite momentum frame of the nucleus, obeying
\be\label{dens}
\int \ d^2 x \ d x_- \ \rho ({\underline x},x_-) = \ A.
\ee
Here we have used the notation of \cite{meM} to describe the
non-Abelian Weizs\"{a}cker-Williams field. Feynman diagrams
corresponding to this classical field were found in \cite{me2} and are
depicted in \fig{wwfig}. As was discussed in some detail in \cite{me2}
the multiple rescatterings are included in the non-Abelian
Weizs\"{a}cker-Williams field in the form of gauge rotations. In
\fig{wwfig} the field of the rightmost nucleon effectively gets rotated 
by the fields of all the other nucleons in the nucleus (see
\fig{wwfig}). As was argued in \cite{me2} classical approximation
corresponds to the limit of no more than two gluons interacting with
each nucleon, which formally means resummation of all powers of the
parameter $\as^2 A^{1/3}$ with $A$ the atomic number of the
nucleus. Each additional power of $\as^2 A^{1/3}$ corresponds to an
extra rescattering and resummation of all of such terms corresponds to
resummation of multiple rescatterings.  For a large nucleus with
$\as^2 A^{1/3} \, \sim \, 1$ all multiple rescatterings are important.

The unintegrated gluon distribution function of a nucleus is
proportional to Fourier transform of the correlator of two
Weizs\"{a}cker-Williams fields in the nuclear wave function, which was
calculated in \cite{jklw,meM} and is given by \eq{wwcorr} below.  This
classical unintegrated gluon distribution has a very peculiar
features: at large values of transverse momentum it falls off like
$1/k_\perp^2$, which is a usual perturbative result. As the transverse
momentum becomes smaller the distribution increases. However below
certain momentum scale, which is called the saturation scale and is
given below by \eq{qs}, the growth of the classical gluon distribution
slows down to just a logarithmic increase, proportional to $\ln
Q_s/k_\perp$. This suggests that multiple rescatterings may help us to
avoid the singularities of collinear factorization by introducing the
saturation scale $Q_s$, which effectively regulates the parton
distributions in the soft momentum region. In terms of this saturation
scale parameter resummation of multiple rescatterings corresponds to
resummation of powers of $Q_s^2/k^2_\perp$, since $Q_s^2 \, \sim \,
\as^2 A^{1/3}$ (see \eq{qs}). The scale determining the value of the
strong coupling constant $\as$ is also the saturation momentum
$Q_s$. Therefore applicability of perturbative QCD and quasi-classical
physics depends strongly on how big $Q_s$ is. If $Q_s^2 \, \gg \,
\Lambda_{QCD}^2$ then $\as (Q_s^2) \, \ll \, 1$ and the physics
described above is applicable to the nuclear scattering process.

As the energy increases (or, equivalently, as we go towards smaller
values of $x$) the quantum corrections become important. Multiple
rescatterings enhanced by quantum corrections become multiple pomeron
exchanges.  There were developed several techniques which resum
multiple pomeron exchanges. The techniques based on resummation of
successive classical emissions in the framework of effective
lagrangian of McLerran-Venugopalan model led to a renormalization
group functional differential equation of \cite{JKLW}. There are
different effective lagrangian approaches, such as the one developed
by Lipatov and collaborators \cite{lks} and another one by Balitsky
\cite{bal}. Finally there is an integral evolution equation which was
obtained in \cite{mme} using the techniques of Mueller's dipole model
\cite{dip}. The equation is similar to the GLR equation of \cite{GLR}
and was also obtained in \cite{bal} for the evolution of the Wilson
lines' correlators. The approach of \cite{JKLW} also seems to be
converging to the equation of \cite{mme} (see
\cite{cons}). Nevertheless, as was argued in \cite{M4,mv} the net
effect of quantum corrections is to increase the saturation scale
$Q_s$ without significantly changing the shape and main qualitative
features of the classical gluon distribution.

\begin{figure}
\begin{center}
\epsfxsize=7cm
\leavevmode
\hbox{ \epsffile{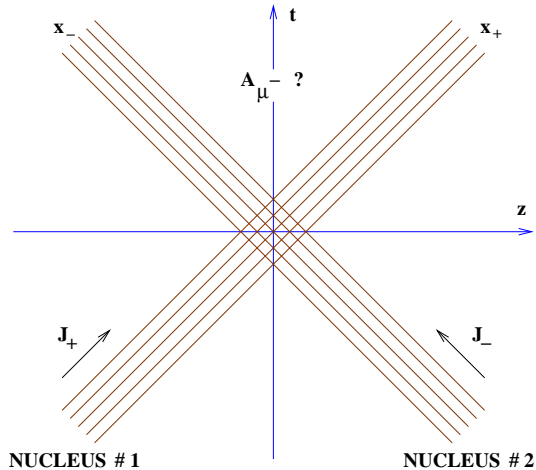}}
\end{center}
\caption{A collision of two ultrarelativistic nuclei at high energies.}
\label{coll}
\end{figure}

But what does the saturation physics teach us about the initial
conditions for the heavy ion collisions? Similarly to how it was done
for the case of a single nucleus structure functions one has to start
by considering purely classical case, neglecting the quantum
corrections. This of course corresponds to not extremely high energy
scattering, $\as \ln s \, \ll \, 1$. For realistic high energies one
has to include quantum evolution corrections, but classical initial
conditions (in the QCD evolution sense) are necessary in order to do
so. Quantum corrections may again only modify the saturation scale in
the classical distribution without significantly changing its shape,
as it happened for nuclear structure functions \cite{M4,mv}. The
problem of inclusion of quantum corrections still has not been
solved. However, since the quantum evolution can be represented as a
series of classical emissions \cite{JKLW,dip} the qualitative picture
of particle production has been constructed in \cite{klm} and certain
physical consequences, such as multiplicity correlations in rapidity
in the produced particle spectrum have been predicted \cite{klm}.

The classical gluon production problem for nuclear or hadronic
collisions has been formulated in \cite{kmw} by Kovner, McLerran and
Weigert. They consider scattering of two ultrarelativistic nuclei from
McLerran-Venugopalan model (see \fig{coll}). The valence quarks of the
nuclei just pass through each other during the collision without
deflection from their straight line light cone
trajectories. Corrections to this approximation are proportional to
positive powers of $x$, and $x \, \ll \, 1$. However, the gluonic
field generated by the collision has non-zero field strength in the
forward light cone and contributes to the gluon production.  If we
would like to obtain an expression for the distribution of produced
gluon which includes all powers of $\as^2 A^{1/3}$ (or, equivalently,
$Q_s/k_\perp$), we can do it by solving classical Yang-Mills equation
with the ultrarelativistic nuclei providing us with the source current
\cite{kmw}. The gluon field given by the solution of the classical
equations of motion in the forward light cone would describe the gluon
production (see \fig{coll}), and, consequently, the initial conditions
for heavy ion collisions. This is the problem we are going to address
in this paper.

The classical field of two nuclei in the forward light cone has been
found in the usual perturbation theory to order $g^3$ in
\cite{kmw,med,mg}. The answer for the distribution of produced gluons 
is proportional to $Q_{s1}^2 Q_{s2}^2 / k_\perp^4$, where $Q_{s1}$ and
$Q_{s2}$ are the saturation scales of the colliding nuclei. This is
the first (lowest order) term of the expansion in powers of
$Q_{s1,2}^2 / k_\perp^2$ and corresponds in this sense to
proton-proton scattering. The answer agrees with the production rate
one would get from employing the so-called Lipatov effective vertex
\cite{BFKL,GLR}, or, equivalently, with the result of Gunion and Bertsch 
\cite{gb}.

A gluon production cross section for a slightly more complicated case
of proton-nucleus interactions was derived in \cite{meM}. In the
formal language of the saturation scale parameters $Q_{s1}$ and
$Q_{s2}$ that cross section includes all powers of $Q_{s1}^2 /
k_\perp^2$ keeping only the leading power of $Q_{s2}^2 / k_\perp^2$.

The problem of finding the classical gluon field in nucleus-nucleus
collisions has been addressed in lattice simulations of Krasnitz and
Venugopalan \cite{kv}. The numerical distribution of the produced
particles has been found and exhibited saturation properties expected,
such as finiteness in the small transverse momentum limit. Mueller
\cite{Mueller2} suggested that the total number of the produced gluons has to 
be proportional to the total number of gluons in the nuclear wave
function with the proportionality coefficient $c$ of order one. The
numerical value of the coefficient was determined in \cite{kv} to be
$c = 1.29 \pm 0.09$.

In this paper we are going to write down an analytical expression for
the distribution of produced gluons which resums all powers of both
$Q_{s1}^2 / k_\perp^2$ and $Q_{s2}^2 / k_\perp^2$. That is we are
going to address the problem of nucleus-nucleus scattering
analytically. The paper is organized as follows: in Sect. II we will
formulate the problem of classical gluon production and make some
useful observations revealing the advantages and disadvantages of
viewing the process in different gauges. In Sect.  IIIA we will review
the solution of the proton-nucleus (pA) problem in covariant gauge
which was given in \cite{meM}, showing how multiple {\it final} state
rescatterings play crucial role in gluon production. In Sect. IIIB we
will analyze the same process of gluon production in pA collisions in
light cone gauge and demonstrate that an entirely different set of
{\it initial} state interactions is important there. We will outline
certain important cancellations of diagrams, which would allow us to
write down an expression for the produced gluons' distribution in
nucleus-nucleus (AA) collisions in Sect. IV. This distribution is
given by \eq{aasol} and is the central result of this paper. It gives
the distribution of gluons in the state immediately following a heavy
ion collision providing initial condition for possible thermalization
of the gluonic system at later times. We explore the properties of the
distribution of \eq{aasol} in Sect. V. There we first obtain a
simplified expression for the distribution in the case of not very
large transverse momenta $k_\perp \, \lsim \, Q_s$, which is given by
\eq{dist2}. We demonstrate that the distribution of produced gluons
(\ref{aasol}) is finite up to $\ln Q_s/k_\perp$ in the soft transverse
momentum limit and scales as $1/k_\perp^4$ when the transverse
momentum gets large. We calculate the typical momentum of the produced
gluons and find that for the case of identical nuclei it is of the
order of the saturation scale $Q_s$, as was conjectured by Mueller in
\cite{Mueller2}. Finally we estimate the coefficient $c$ from
\cite{Mueller2} and find $c \, \approx \, 2 \ln 2$, which is close to 
the result of \cite{kv}. In Sect. VI we estimate the saturation scale
using the new data from PHOBOS experiment at RHIC \cite{pho} and
obtain $Q_s^2 \, \approx \, 2 \, \mbox{GeV}^2$, which is marginally in
the perturbative QCD region and agrees with the result of
\cite{ekrt}. We end the paper with a summary of the results obtained.

\section{Formulation of the Problem}

In this section we are going to review the formulation of the problem
of finding the classical gluon field of two colliding nuclei and make
some observations which will be useful later.

As was originally stated in \cite{kmw} one needs to solve the
classical Yang--Mills equations
\be\label{ym}
D_\mu F^{\mu \nu} = J^{\nu}\,\, ,
\ee
with the current $J^{\mu}$ arising due to the valence quarks in the
colliding nuclei. The valence quarks move ultra-relativistically along
the straight lines on the light cone and do not get deflected in the
collision. (Deflection is suppressed by a power of the center of mass
energy of the colliding system.) Thus the current generated by them
has non-zero light cone components $J_+$ and $J_-$ and zero transverse
component ${\underline J} \, = \, 0 $. In a non-Abelian theory the
source current is not gauge invariant, it gets rotated under gauge
transformations. We will first construct the current in a particular
gauge --- covariant gauge
\be
\partial_\mu \, A_\mu \, = \, 0.
\ee
Before the collision the nuclei do not see each other and do not
interact. The current is given by the sum of the currents of two free
nuclei on the light cone. As was shown in \cite{me,med} the current of
a free nucleus in covariant gauge is given by a simple superposition
of the currents of the point color charges (valence quarks) in the
nucleus, each of them being parametrically of order $g$ in strong
coupling constant. For example, in the model of quarkonium nucleus
considered in \cite{me,med}, where the nucleus was envisaged as an
ensemble of point color charges, with each nucleon consisting of two
valence quarks (a quark and an antiquark), the free current is
\begin{mathletters} \label{current0}
\begin{eqnarray} 
J_+^{cov \ (0)} & = & g \sum_{i=1}^{A_1} T^a \, (T_i^a)\, [\delta(x_- -
x_{i-})\, \delta( {\underline x}-{\underline x}_i ) - \delta(x_- -
x'_{i-})\, \delta( {\underline x}-{\underline x}'_i ) ]\,\, , \\ 
J_-^{cov \ (0)}
& = & g \sum_{j=1}^{A_2} T^a \, (\tilde{T}_j^a)\, [\delta(x_+ -
y_{j+})\, \delta( {\underline x}-{\underline y}_j ) - \delta(x_+ -
y'_{j+})\, \delta( {\underline x}-{\underline y}'_j )]\,\, , \\ 
{\underline J}^{cov \ (0)} & = & 0\,\, .
\end{eqnarray}
\end{mathletters}
Here $x_i$'s ($x'_i$'s) and $y_j$'s ($y'_j$'s) are positions of the
quarks (antiquarks) in the $i$th nucleon of the first and in the $j$th
nucleon of the second nucleus correspondingly. The first nucleus is
moving in the ``+'' direction, while the second is moving in the
``$-$'' direction (\fig{coll}). $(T_i^a)$ and $(\tilde{T}_j^a)$ are
$SU(3)$ color generators of the $i$th nucleon in the first nucleus and
of the $j$th nucleon in the second nucleus. $A_1$ and $A_2$ are the
atomic numbers of the nuclei.

\eq{ym} implies conservation of the classical current $J_\mu$
\be\label{ccons}
D_\mu \, J_\mu \, = \, 0.
\ee
Similarly to what was done in \cite{med} we will use this condition to
construct the covariant gauge current $J^{cov}_\mu$ during and after
the collision. The valence quarks do not get deflected from their
light cone trajectories in the collision. The only non-zero components
of the current are therefore $J_+$ and $J_-$. We can rewrite
\eq{ccons} as
\be\label{ccons1}
\partial_+ \, J^{cov}_- +  \partial_- \, J^{cov}_+ - i g [A^{cov}_+ , J^{cov}_- ] 
- i g [A^{cov}_-, J^{cov}_+] = 0. 
\ee
\eq{ccons1} can be satisfied with the following ansatz for the current
\be\label{covcur}
J^{cov}_+ \, = \, U^{-1} (x) \, J^{cov \ (0)}_+ (x) \, U (x), \hspace*{1cm}
J^{cov}_- \, = \, S^{-1} (x) \, J^{cov \ (0)}_- (x) \, S (x),
\ee
where
\be\label{u}
U (x) \, = \, \mbox{P} \exp \left( - i g \int_{-\infty}^{x_+} \ d x'_+ \
A^{cov}_- \right)
\ee
and
\be\label{s}
S (x) \, = \, \mbox{P} \exp \left( - i g \int_{-\infty}^{x_-} \ d x'_- \
A^{cov}_+ \right),
\ee
with $A^{cov}_\pm$ the components of the unknown solution of
\eq{ym}. Thus generally speaking the matrices $U (x)$ and $S (x)$ are
not known. In arriving at the solution of \eq{ccons1} given by
Eqs. (\ref{covcur}), (\ref{u}) and (\ref{s}) we have fixed the initial
conditions for \eq{ccons1}: we required that before the collision the
current $J^{cov}_\mu$ should be given by the free nuclear current
$J^{cov \ (0)}_\mu$ of \eq{current0}. This is just a casuality
requirement which makes sure that there is no interactions between the
nuclei prior to collision.  Eqs. (\ref{covcur}), (\ref{u}) and
(\ref{s}) certainly satisfy this initial condition, since before the
collision $U \ = \ S \ = \ 1$, as the fields of free nuclei are
non-zero only on light cone (see for instance
\cite{me,Mueller1}). Since \eq{ccons1} is a linear differential
equation the solution of \eq{covcur} is unique for the given initial
condition of \eq{current0}.

\eq{covcur} has a very simple physical interpretation: the 
valence quarks in the nuclei during and after the collision are still
moving along the same straight lines on the light cone. The only
effect of the collision on these valence quarks is the rotation of
their color charges by the gluon field created in the collision and by
the gluon field of the other nucleus. This has been discussed and
illustrated in \cite{med} at the lowest nontrivial order in $\as$.

Eqs. (\ref{ym}) and (\ref{covcur}) provide us with complete
formulation of the problem in covariant gauge. We have to solve the
Yang-Mills equations (\ref{ym}) with the conserved current
(\ref{covcur}). We are now going to demonstrate an interesting
property of the current.

Let us perform a gauge transformation with the matrix $S (x)$.  The
new gluon field will be given by
\be
A^{LC}_\mu \, = \, S \ A^{cov}_\mu S^{-1} \, - \, \frac{i}{g} \
(\partial_\mu S) \ S^{-1}.
\ee
As easy to see $A^{LC}_+ \, = \, 0$, which means that the gauge
transformation with the matrix $S (x)$ transforms the field into the
light cone gauge. The current in the light cone gauge is
\be\label{lccur}
J^{LC}_+ \, = S (x) \, U^{-1} (x) \, J^{cov \ (0)}_+ (x) \, U (x) \,
S^{-1} (x) , \hspace*{1cm} J^{LC}_- \, = \, J^{cov \ (0)}_- (x) .
\ee
From \eq{lccur} it follows that the ``$-$'' component of the current
in the $A^{LC}_+ \, = \, 0$ light cone gauge remains unchanged
throughout the collision and is equal to the order $g$ ``initial''
free nucleus current in the covariant gauge. That means that the
charges of the second nucleus in the light cone gauge do not get
rotated in the collision. Let us illustrate what this statement means
in terms of diagrams.

\begin{figure}
\begin{center}
\epsfxsize=15cm
\leavevmode
\hbox{ \epsffile{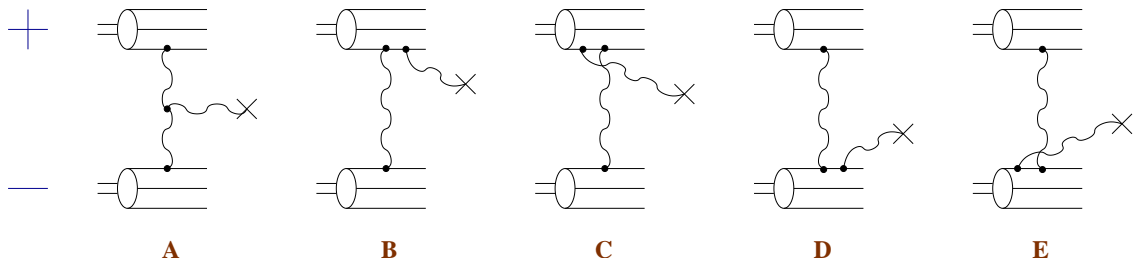}}
\end{center}
\caption{Diagrams contributing to the classical gluon field in covariant 
gauge at order $g^3$.}
\label{lip}
\end{figure}

In \cite{kmw,med,mg} the classical gluon field of two colliding nuclei
was found perturbatively at the lowest non-trivial order in $g$, which
happened to be order $g^3$. The diagrams contributing to the gluon
field at this order in covariant gauge are shown in
Fig. \ref{lip}. There we present a collision of two ultrarelativistic
nucleons, the upper one of which is moving in the light cone ``plus''
direction while the lower one is moving in the ``minus''
direction. The straight lines in \fig{lip} correspond to the valence
quarks. The cross denotes the point in coordinate space where one
measures the field.

The interpretation of the diagrams of Fig. \ref{lip} has been given in
\cite{med}. In diagram A two gluon fields merge to produce the final field.  
In the diagrams B and C the current of the upper quark gets rotated by
the field of the lower quark and a gluon is emitted off the modified
current. In the diagrams D and E the opposite happens: the current of
the lower quark gets rotated by the field of the upper quark and
emits a gluon. In both cases of B,C and D,E the current of one of the
quarks gets a rotational correction of the order $g^2$ (one gluon
exchange contribution) which could also be obtained from \eq{covcur}
by perturbative expansion of the matrices $U (x)$ and $S (x)$ to the
lowest non-trivial order \cite{med}. 

\begin{figure}
\begin{center}
\epsfxsize=10cm
\leavevmode
\hbox{ \epsffile{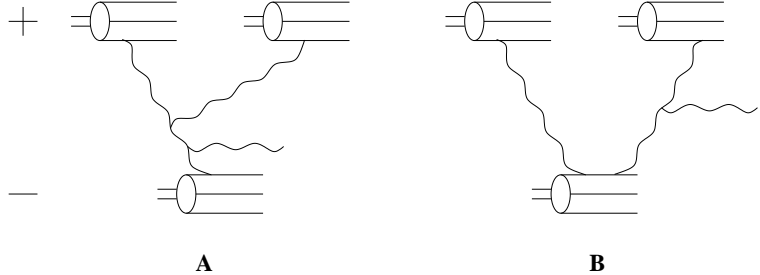}}
\end{center}
\caption{$A_+ = 0$ light cone gauge diagrams: A contributes to the classical gluon 
field while B does not.}
\label{lcexp}
\end{figure}

Now looking at the light cone gauge current given by \eq{lccur} we
see that $J^{LC}_\mu$ does not get any rotational corrections
whatsoever and remains at the lowest order in $g$ (free nucleus
current). Therefore we can conclude that, for instance, diagrams D and
E of \fig{lip} do not contribute to the classical gluon field in $A_+
\ = \ 0$ light cone gauge, which is a self-evident statement. However, 
we can draw more general and less trivial conclusions from
\eq{lccur}. Since the current of the second nucleus does not get
rotated it means that it only rotates the current and/or the field of
another nucleus. This also implies that there is no diagrams with more
than one gluon line connecting to any quark line in the second nucleus
(except for virtual diagrams where two gluons can connect to a single
quark line -- see \cite{me2}). The statement is illustrated in
\fig{lcexp}. Diagram in \fig{lcexp}A has only one gluon line attaching
to the quark line in the second (lower) nucleus moving in the
``minus'' direction and therefore may contribute to the classical
field. The diagram of \fig{lcexp}B has two gluons interacting with the
quark in a nucleon of the second nucleus, and the nucleon remains in a
color non-neutral state at the end. This diagram does not contribute
to the classical field according to what we have shown in
\eq{lccur}.

\section{Proton-Nucleus Collisions Revisited}

Before addressing the issue of nucleus--nucleus collisions (AA) let us
first consider a somewhat easier problem of proton--nucleus collisions
(pA). Below we are going to review the solution of the problem in
covariant gauge given in \cite{meM} and then proceed by analyzing the
same pA process in the light cone gauge of the nucleus.  Our
interpretation of the underlying light cone gauge physics will be
slightly different from the one presented in \cite{meM}.

\subsection{Covariant Gauge}

\begin{figure}
\begin{center}
\epsfxsize=15cm
\leavevmode
\hbox{ \epsffile{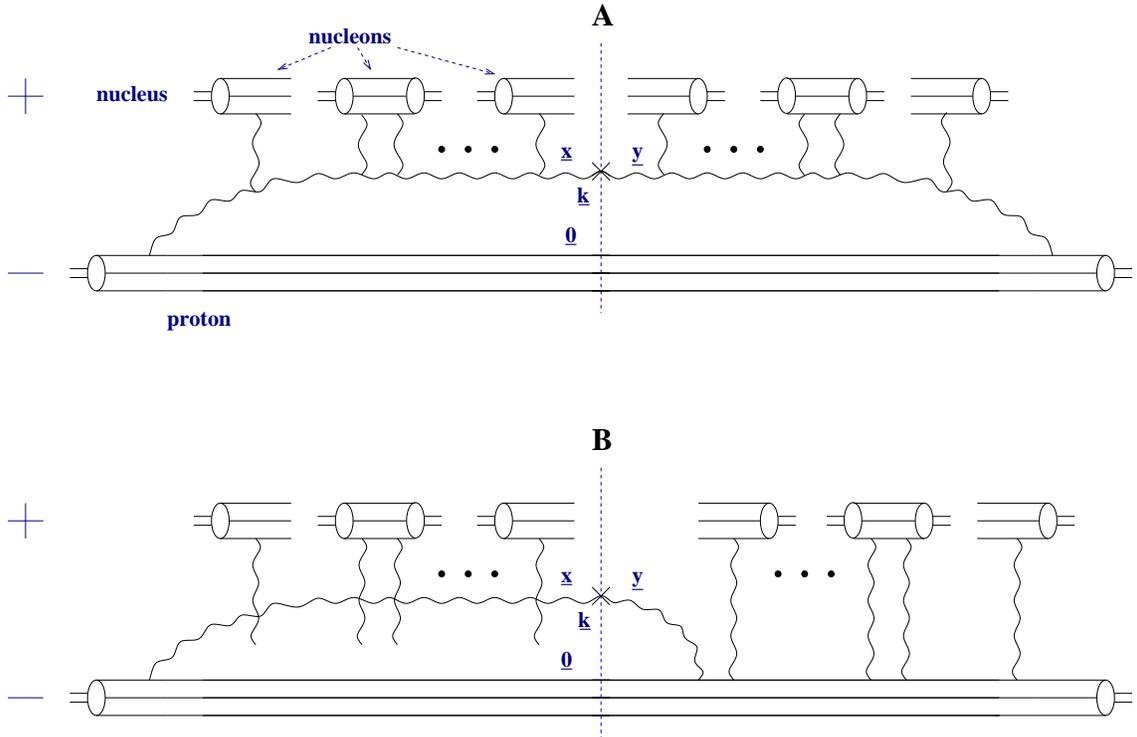}}
\end{center}
\caption{Covariant gauge (or more accurately $A_- = 0$ gauge) gluon 
production diagrams for proton--nucleus collision as considered in
\protect\cite{meM}. Multiple rescatterings in the nucleus determine the 
interactions in this gauge.}
\label{pafig}
\end{figure}

Consider a collision of an ultrarelativistic nucleus moving in the
``plus'' light cone direction and a proton moving in the ``minus''
light cone direction. For pA collisions we want to solve the same
problem of classical gluon production as was stated above for AA
collisions. In this subsection we will just follow the discussion of
\cite{meM}. We will work in $A_- \ = \ 0$ gauge with polarization vector 
also taken in that gauge, $\epsilon_- \ = \ 0$, which for the nucleus
moving in the ``plus'' direction is equivalent to covariant gauge
($\partial \cdot A \ = \ 0$) \cite{meM}. Following \cite{meM} we will
consider the process in the rest frame of the nucleus and perform the
calculations in the light cone perturbation theory (see \cite{bl} and
references therein). Then the physical picture of the gluon production
is the following: the incoming proton may already have a gluon in its
light cone wave function before the collision with the nucleus and the
system of the proton and gluon multiply rescatters on the nucleons in
the nucleus. Alternatively the proton can emit the gluon after the
multiple rescatterings in the nucleus. The diagrams where the gluon is
emitted during the proton's passing through the nucleus are suppressed
by powers of its large light cone momentum $p_-$, i.e., by powers of
center of mass energy of the system (eikonal approximation)
\cite{meM}. Multiple rescatterings are easier to resum by calculating 
the amplitude in the transverse coordinate space \cite{bdmps,meM}. To
obtain the gluon production cross section we have transform the
amplitude into the momentum space and square it. The diagrams
contributing to the gluon production cross section are shown in
\fig{pafig}. The graph in \fig{pafig}A corresponds to the square of
the amplitude corresponding to the case when the gluon is present in
proton's wave function before the collision. The diagram in
\fig{pafig}B gives the interference term between the amplitude from
\fig{pafig}A and the amplitude in which the gluon is emitted by the proton 
after the collision. Of course a diagram complex conjugate to
\fig{pafig}B should also be included. It can be shown that the square of 
the diagram with late gluon emission does not have any interactions in
it and can be neglected. (The gluon exchanges between the proton and
the nucleus cancel.) In \cite{meM} the interactions counting was a
little different from the one we will present below and that diagram
was included, leading to the same result.

Note also that in the quasi--classical approximation depicted in
\fig{pafig} the interaction is modeled by single and double gluon
exchanges. The limit of no more than two gluons per nucleon is imposed
\cite{me2}. If a particular nucleon exchanges a gluon with the rest of
the system in the amplitude then it has to exchange a gluon in the
complex conjugate amplitude to remain color neutral. Alternatively the
nucleon can exchange two gluon in the amplitude (complex conjugate
amplitude) , but then it can not interact in the complex conjugate
amplitude (amplitude). This is done in the spirit of the
quasi--classical approximation resumming all powers of $\as^2 \
A^{1/3}$, as was discussed in the Introduction.

In the graph of \fig{pafig}A the nucleons of the nucleus interact with
both the proton and the gluon by gluon exchanges. It was noticed in
\cite{meM} that the interactions with the proton can be neglected due
to real--virtual cancellation. Moving a gluon exchanged between the
nucleus and the proton across the cut does not change the momentum of
the produced gluon in \fig{pafig}A but does change the sign of the
whole term, causing the cancellation. That is why we have to consider
only the interactions with the gluon in \fig{pafig}A.  Similar kind of
cancellation does {\it not} happen in \fig{pafig}B. Moving an
exchanged (Coulomb) gluon across the cut would force us to move it
across the gluon emission vertex for the produced gluon on the right
hand side, thus changing the momentum of the produced gluon. Thus all
the possible interactions have to be included in \fig{pafig}B. On the
right hand side of the diagram in \fig{pafig}B only the interactions
with the proton are possible.

To obtain the answer for the gluon production cross section in pA in
the quasi--classical approximation we have to convolute the wave
function of the proton with a soft gluon in it with the Glauber-type
propagator. The answer has been derived in \cite{meM} (see also
\cite{kst} and references therein). The diagram in \fig{pafig}A gives \cite{meM}
\begin{mathletters} \label{pasol}
\be
\frac{d \sigma^{pA}_1}{d^2 k \ dy} \ = \ \frac{1}{\pi} \, \int \ d^2 b \, d^2 x 
\, d^2 y \, \frac{1}{(2 \pi)^2} \, \frac{\as C_F}{\pi} \frac{{\underline x} 
\cdot {\underline y}}{{\underline x}^2 {\underline y}^2} \, e^{i {\underline k} 
\cdot ({\underline x} - {\underline y})} \, \left( e^{- ({\underline x} - 
{\underline y})^2 \ Q_s^2 /4 } - 1 \right)
\ee
and the diagram in \fig{pafig}B plus its complex conjugate after a
somewhat more sophicticated calculation \cite{meM} gives
\cite{meM}
\be
\frac{d \sigma^{pA}_{2+3}}{d^2 k \ dy} \ = \ \frac{1}{\pi} \, \int \ d^2 b \, d^2 x 
\, d^2 y \, \frac{1}{(2 \pi)^2} \, \frac{\as C_F}{\pi} \frac{{\underline x} 
\cdot {\underline y}}{{\underline x}^2 {\underline y}^2} \, e^{i {\underline k} 
\cdot ({\underline x} - {\underline y})} \, \left( 1 -  e^{- {\underline x}^2 \ 
Q_s^2 /4 } + 1 -  e^{- {\underline y}^2 \ Q_s^2 /4 } \right).
\ee
\end{mathletters}
The total gluon production cross section is equal to the sum of the
terms in \eq{pasol}
\be
\frac{d \sigma^{pA}}{d^2 k \ dy} \ = \ \frac{d \sigma^{pA}_1}{d^2 k \ dy} 
+ \frac{d \sigma^{pA}_{2+3}}{d^2 k \ dy}.
\ee
In \eq{pasol} ${\underline x}$ and ${\underline y}$ are the transverse
coordinates of the gluon in the amplitude and the complex conjugate
amplitude correspondingly counted with respect to the transverse
position of the quark in the proton off which the gluon is emitted
(${\underline 0}$). $b$ is the impact parameter. ${\underline k}$ is
the gluon's transverse momentum. Same as in \cite{meM} we use a
shorthand notation (see also
\cite{bdmps})
\be\label{xqs}
{\underline x}^2 Q_s^2 \ = \ {\underline x}^2 \ \frac{8 \pi^2 \as N_c
R}{N^2_c - 1} \, \rho \, xG (x, 1/{\underline x}^2),
\ee
with $\rho$ the density of \eq{dens} taken in the nuclear rest frame.
In the two gluon approximating the gluon distribution function of a
nucleon is
\be\label{xg}
xG (x, 1/{\underline x}^2) \ = \ \frac{\as C_F}{\pi} \, \ln
\frac{1}{{\underline x}^2 \mu^2},
\ee
with $\mu$ some infrared cutoff. For simplicity of calculations
throughout the paper we assume that the nucleus has a cylindrical
shape, with radius $R$ and the height of the cylinder $2 R$. The
cylinder is lined up along the $z$ axis. Generalization of our results
to a spherical nucleus is trivial.

In general the saturation scale $Q_s^2$ has to be found from the
following implicit equation \cite{meM,bdmps}
\be\label{qs}
Q_s^2 \ = \ \frac{8 \pi^2 \as N_c R}{N^2_c - 1} \, \rho \, xG (x,
Q_s^2).
\ee
However, since the logarithm in \eq{xg} is a slowly varying function
we can assume that in our classical approximation without any QCD
evolution in the structure functions the gluon distribution function
is approximately a constant and the right hand side of \eq{qs} is
independent of $Q_s$. Thus \eq{qs} turns from implicit equation into
an equality.

To summarize the results reviewed here we note that the gluon
production in pA collisions in covariant gauge is driven by the
multiple final state interactions. Now we will explore how this
picture changes in the light cone gauge of the nucleus.

\subsection{Light Cone Gauge}

Let us now consider pA scattering in the $A_+ = 0$ light cone
gauge with polarization vector also taken in the same gauge
$\epsilon_+ \ = \ 0$. A direct analysis of the light cone gauge
diagrams could be a little difficult \cite{meM}. We are going to use a
different strategy, which was already employed previously in
\cite{me2}. We know the answer for the gluon production cross section 
given by \eq{pasol}. Here we are going to guess the diagrams in the
$A_+ = 0$ light cone gauge which give us the same answer. Once we
guessed the correct diagrams we can conclude that the remaining
diagrams should cancel with each other, since they do not contribute
to the cross section.

\begin{figure}
\begin{center}
\epsfxsize=15cm
\leavevmode
\hbox{ \epsffile{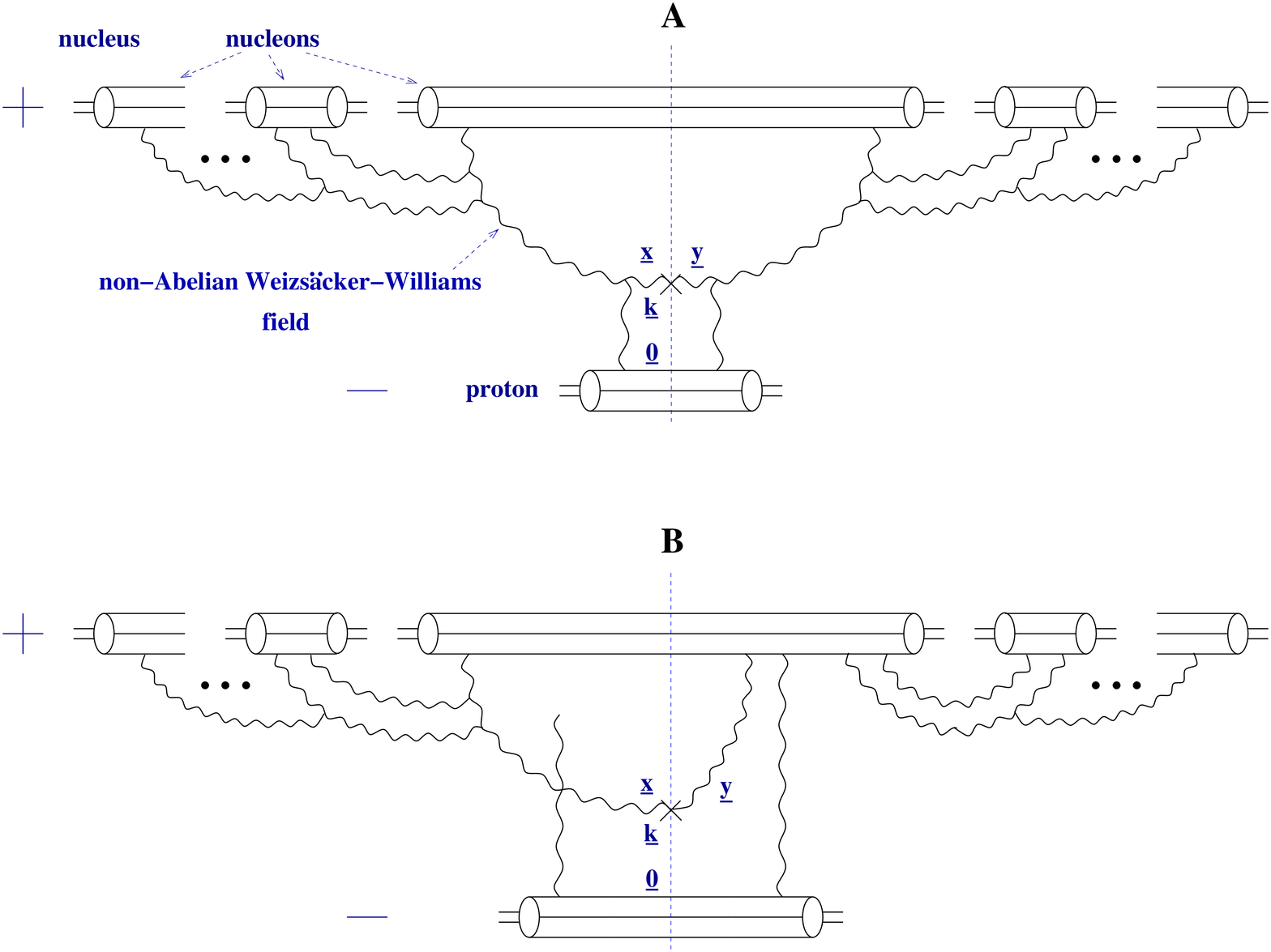}}
\end{center}
\caption{Gluon production in proton--nucleus collisions in $A_+ = 0$ light 
cone gauge (see text).}
\label{palc}
\end{figure}

The light cone gauge diagrams contributing to the gluon production
cross section in proton--nucleus collisions are depicted in
\fig{palc}. The pA scattering process could be viewed in either rest 
frame of the proton or in the center of mass frame. Again we are going
to perform the calculation in the framework of the light cone
perturbation theory. Similar to the covariant gauge case considered
above the incoming nucleus can emit a gluon in its wave function
either before or after the collision with the proton. The one gluon
light cone wave function of an ultrarelativistic nucleus is given by
${\underline A}^{WW} ({\underline x}) \cdot {\underline\epsilon}$,
with ${\underline A}^{WW}$ the non-Abelian Weizs\"{a}cker-Williams
field of the nucleus given by \eq{ww} with suppressed $x_-$ dependence
\cite{me,jklw} and ${\underline\epsilon}$ the polarization vector in
light cone gauge. (One can see that it is the case for instance by
calculating the one gluon rescattering diagram on the left hand side
of \fig{palc}A with this wave function and by using conventional
perturbation theory. Both results give the same answer.)
Diagrammatically the light cone wave function corresponds to the same
set of diagrams as was depicted above in \fig{wwfig}. The fields of
the nucleons in the nucleus ``gauge rotate'' the
Weizs\"{a}cker-Williams field of one of the nucleons \cite{me2}. The
interaction with the proton can only be by the means of single or
double gluon exchanges, as was shown in Sect. II. \eq{lccur} shows
that the ``minus'' component of the current does not get rotated
implying that there could not be more than two gluons exchanged with
the proton in the $A_+ = 0$ light cone gauge.

Before colliding with the proton the nucleus can develop the
Weizs\"{a}cker-Williams one gluon light cone wave function which then
interacts with the proton by means of one or two gluon exchanges,
according to the rules of the quasi--classical approximation
\cite{me2,meM}.  The square of the graph corresponding to this scenario
is shown in \fig{palc}A. As in \fig{pafig} the interactions of the
proton with the nucleons in the nucleus cancel through the
real--virtual cancellation leaving only the interactions with the
gluon line. One may notice that the final state interactions are left
out in the diagram of \fig{palc}A, but as we will show below we do not
need them to reproduce the contribution of the graph in \fig{pafig}A,
which implies that they cancel with each other.

The second possible scenario corresponds to the case when there is no
gluon in the nuclear wave function by the time the collision happens
and the gluon is emitted by the nucleus after the interaction with the
proton. Then the nuclear wave function without an emitted gluon
corresponds to the fields of the nucleons rotating the current of one
of the nucleons in the nucleus. This is shown on the right hand side
of \fig{palc}B. The nucleon then interacts with the proton by
exchanging one or two gluons with it. After that the nucleus can emit
a gluon to be produced in the final state. Another possibility which
is not shown in \fig{palc}B but which contributes to the gluon
production corresponds to the case when the Weizs\"{a}cker-Williams
gluon is present in the nuclear wave function by the time of the
collision, similar to \fig{palc}A, but after the interaction with the
proton the gluon merges into the quark line of one of the nucleons,
which later re-emits the gluon. We could not find an {\it a priori}
argument prohibiting an emission of the whole Weizs\"{a}cker-Williams
field after the interaction. However, as we will see below one needs
to emit only one gluon to be able to reproduce the results of the
previous section. The square of the diagram on the right hand side of
\fig{palc}B is zero since the interactions cancel due to real--virtual
cancellation \cite{meM}. The only contribution we get from it is the
interference term depicted in \fig{palc}B. There on the left hand side
we have the same diagram as in \fig{palc}A except that now
interactions of the proton with the ``last'' nucleon in the nucleus do
not cancel. We will show that the diagram of \fig{palc}B provides us
with the contribution equal to that of the graph in \fig{pafig}B.

Let us now calculate the diagrams in \fig{palc}. The contribution of
\fig{palc}A can be obtained by convoluting the correlation function of
the fields on both sides of the cut with the gluon--proton
interactions amplitude. The result yields
\be\label{lc11}
\frac{d \sigma^{pA}_{LC1}}{d^2 k \ dy} \ = \ \int \frac{d^2 x \ 
d^2 y}{(2 \pi)^2} \, e^{i {\underline k} \cdot ({\underline x} -
{\underline y})}\, \frac{2}{\pi} \, \mbox{Tr} \, \left< {\underline A}^{WW}
({\underline x}) \cdot {\underline A}^{WW} ({\underline y}) \right> \,
\frac{- \as \pi^2 N_c}{N_c^2 - 1} \, ({\underline x} - {\underline y})^2
\, xG (x, 1/({\underline x} - {\underline y})^2).
\ee
In \cite{jklw,meM} the correlation function of two non-Abelian
Weizs\"{a}cker-Williams fields in the nuclear wave function was found
to be
\be\label{wwcorr}
\mbox{Tr} \left< {\underline A}^{WW} ({\underline x}) \cdot {\underline A}^{WW} 
({\underline y}) \right> \, = \, \frac{C_F}{\pi \as ({\underline x} -
{\underline y})^2} \, \left(1 - e^{ - ({\underline x} - {\underline
y})^2 \, Q_s^2 / 4 } \right).
\ee
Employing Eqs. (\ref{wwcorr}) and (\ref{xg}) in \eq{lc11} and defining
new variables ${\underline z} = {\underline x} - {\underline y}$ and
${\underline b} = {\underline y}$ we obtain
\be\label{lc12}
\frac{d \sigma^{pA}_{LC1}}{d^2 k \ dy} \ = \ \int \, d^2 b \, d^2 z \, 
e^{i {\underline k} \cdot {\underline z}} \, \frac{1}{(2 \pi)^2} \,
\frac{\as C_F}{\pi} \, \ln \frac{1}{{\underline z}^2 \mu^2} \, \left( 
e^{ - {\underline z}^2 \, Q_s^2 / 4 } - 1 \right).
\ee
Using Eqs. (63) and (64) from \cite{meM} one can see that
\be\label{for}
\ln \frac{1}{{\underline z}^2 \mu^2} \, = \, \frac{1}{\pi} \, \int 
d^2 y \, \frac{{\underline y} \cdot ({\underline z} + {\underline
y})}{{\underline y}^2 ({\underline z} + {\underline y})^2}
\ee
where the $y$ integration is cut off by $1/\mu$ in the infrared
limit. Inserting \eq{for} into \eq{lc12} and comparing the result to
Eq. (\ref{pasol}a) one can see that
\be
\frac{d \sigma^{pA}_{LC1}}{d^2 k \ dy} \ = \ \frac{d \sigma^{pA}_{1}}{d^2 k \ dy}.
\ee
Thus we have shown that the contribution of the diagrams in
\fig{palc}A is equal to the contribution of the diagrams in
\fig{pasol}A.

The calculation of the graphs depicted in \fig{palc}B is a little more
complicated. Similar to diagrams of \fig{palc}A a correlator of two
gluonic fields is involved. However, in the field on the right hand
side of the cut in \fig{palc}B only rotations of the source are
allowed. Similarly to what was done in \cite{me2,meM} we argue that
the fields of the nucleons rotate the quark line to which the emitted
final state gluon is attached, as well as the Coulomb gluon's field
coming from the proton below. Everything is gauge rotated, except for
the gluon line of the emitted gluon. After applying Ward identities we
conclude that effectively only the vertex where the emitted gluon
connects to the quark line on the right hand side of \fig{palc}B is
rotated. This corresponds to rotation of the source of the gluon field
emitted \cite{me2}. The effect of this gauge rotation is modification
of the expression for the field from \eq{ww} into
\be\label{wwm}
{\underline A}^{WW}_{mod} ({\underline x},x_-) \ = \ \int \ d^2 z \ d z_-
\theta (z_- - x_-) \ {\hat \rho}^a ({\underline z},z_-) \
\frac{{\underline x} - {\underline z}}{|{\underline x} - {\underline z}|^2} 
\ S_0 ({\underline z},z_-) T^a S_0^{-1} ({\underline z},z_-).
\ee
After we sum over the all possible connections of the Coulomb gluon
lines to the ``last'' nucleon in the nucleus in \fig{palc}B the
resulting expression would depend on the transverse coordinate of the
quark in that nucleon to which the Coulomb gluon couples. Thus we can
not factorize the averaging in the nuclear wave function from the
interaction terms anymore, like it was done in obtaining
\eq{lc11}. With all the above-mentioned complications in mind we write
the contribution of the diagram in \fig{palc}B as (see \cite{meM} for
a similar calculation)
\ben
\frac{d \sigma^{pA}_{LC2}}{d^2 k \ dy} \ = \ \frac{2}{\pi} \, \int 
\frac{d^2 x \ d^2 y}{(2 \pi)^2} \, e^{i {\underline k} \cdot ({\underline x} -
{\underline y})} \, \int d^2 z \, d z_- \, d^2 z' \, d z'_- \, \left<
\frac{{\underline x} - {\underline z}}{|{\underline x} - {\underline
z}|^2} \cdot \frac{{\underline y} - {\underline z'}}{|{\underline y} -
{\underline z'}|^2} \, {\hat \rho}^a ({\underline z},z_-) \, {\hat
\rho}^b ({\underline z}',z'_-) \right.
\een
\be\label{lc21}
\left. \times \mbox{Tr} \left[ S_0 ({\underline x},z_-) 
T^a S_0^{-1} ({\underline x},z_-) S_0 ({\underline z}',z'_-) T^b
S_0^{-1} ({\underline z}',z'_-) \right] \, \frac{\as \pi^2 N_c}{N_c^2
- 1} \, ({\underline x} - {\underline z})^2
\, xG (x, 1/({\underline x} - {\underline z})^2)\right>
\ee
Similar to what was done before in \cite{me,meM} we argue that we can
average the densities ${\hat \rho}^a ({\underline z},z_-) \, {\hat
\rho}^b ({\underline z}',z'_-)$ in the nuclear wave function
independently and we can employ \eq{dcorr} to do so. In \cite{meM} the
following result was derived (see Eq. (48) in \cite{meM})
\ben
\left< \mbox{Tr} \left[ S_0 ({\underline x},z_-) 
T^a S_0^{-1} ({\underline x},z_-) S_0 ({\underline z},z_-) T^a
S_0^{-1} ({\underline z},z_-) \right] \right> \ = 
\een
\be\label{scorr}
= \ C_F N_c \exp \left( - \frac{\as \pi^2 N_c}{N_c^2 - 1} \, ({\underline
x} - {\underline z})^2 \, \rho ({\underline x},z_-) \, xG (x,
1/({\underline x} - {\underline z})^2) \, (z_- + z_{0-})\right),
\ee
with $\pm z_{0-}$ is the upper (lower) limit of the $z_-$ integration
in \eq{lc21}. In \eq{scorr} we assume that $\rho ({\underline x},z_-)$
does not vary much between ${\underline x}$ and ${\underline z}$,
which is justified for a large nucleus. Using Eqs. (\ref{dcorr}) and
(\ref{scorr}) in \eq{lc21} we obtain
\be\label{lc22}
\frac{d \sigma^{pA}_{LC2}}{d^2 k \ dy} \ = \ \frac{1}{\pi} \, \int 
\frac{d^2 x \ d^2 y}{(2 \pi)^2} \, e^{i {\underline k} \cdot ({\underline x} -
{\underline y})} \, d^2 z \, \frac{\as C_F}{\pi} \,
\frac{{\underline x} - {\underline z}}{|{\underline x} - {\underline
z}|^2} \cdot \frac{{\underline y} - {\underline z}}{|{\underline y} -
{\underline z}|^2} \, \left(1 - e^{ - ({\underline x} - {\underline
z})^2 \, Q_s^2 / 4 } \right).
\ee
Defining new variables ${\underline x} = {\underline x} - {\underline
z}$, ${\underline y} = {\underline y} - {\underline z}$ and
${\underline b} = {\underline z}$ and adding the contribution of the
complex conjugate to \fig{palc}B diagram ($ d
\sigma^{pA}_{LC3} / d^2 k \ dy$) we can compare the result with
Eq. (\ref{pasol}b) and conclude that
\be
\frac{d \sigma^{pA}_{LC2}}{d^2 k \ dy} + 
\frac{d \sigma^{pA}_{LC3}}{d^2 k \ dy} \ = \ \frac{d \sigma^{pA}_{2+3}}{d^2 k \ dy}.
\ee
We have thus shown that the contribution of the diagrams in
\fig{palc}B is equal to the contribution of the diagrams in
\fig{pafig}B.

\begin{figure}
\begin{center}
\epsfxsize=15cm
\leavevmode
\hbox{ \epsffile{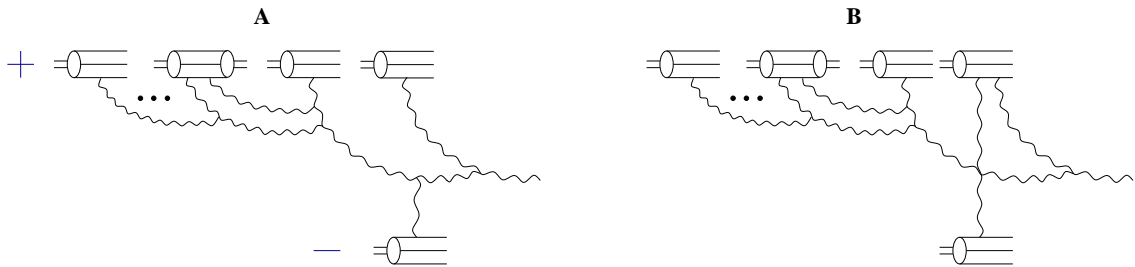}}
\end{center}
\caption{Some of the final state interaction diagrams which do not 
contribute to gluon production in pA collisions in $A_+ = 0$ light 
cone gauge.}
\label{zero}
\end{figure}

We have proved that the diagrams in \fig{palc} are the only diagrams
contributing to the gluon production cross section in $A_+ = 0$ light
cone gauge. Thus a whole class of diagrams with final state
interactions does not contribute to the cross section. Several
examples of such graphs are shown in \fig{zero}.

The diagram in \fig{zero}A represents a class of diagrams where the
non-Abelian Weizs\"{a}cker-Williams wave function of the nucleus after
interaction with the proton merges with another (non-interacting)
Weizs\"{a}cker-Williams wave function. The graph in \fig{zero}B can be
viewed as a similar to \fig{zero}A process, where after proton-nucleon
interaction a gluon field is emitted, which later on merges with the
non-Abelian Weizs\"{a}cker-Williams wave function of the
nucleus. There is another class of diagrams which do not contribute to
the cross section where two Weizs\"{a}cker-Williams wave functions
merge with each other and with a Coulomb gluon coming from the proton
through a four-gluon vertex, producing a gluon in the final
state. Those diagrams are probably suppressed in the old-fashioned
light cone perturbation theory as requiring the gluon fields' merger
to happen at a particular light cone time when the system is passing
the proton.

As was shown above all of the final state interactions shown in
\fig{zero} do not contribute to the gluon production cross section in
pA collisions. Therefore they should cancel, either with each other or
individually due to some other cancellation mechanism. From
considering gluon production in pA collisions in the light cone gauge
we may draw the following conclusion: the gluons produced by the
collision do not merge with the non-Abelian Weizs\"{a}cker-Williams
wave function of the nucleus.

\begin{figure}
\begin{center}
\epsfxsize=9cm
\leavevmode
\hbox{ \epsffile{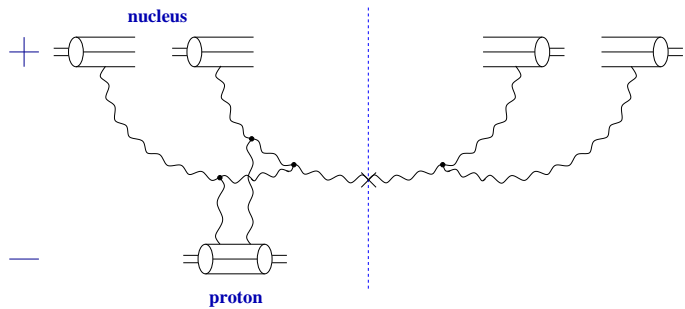}}
\end{center}
\caption{A virtual diagram which does not contribute to the gluon 
production process in pA collisions in $A_+ = 0$ light cone
gauge. Triple gluon vertices are marked with black dots.}
\label{zeropa}
\end{figure}

Another important diagram which {\it a priori} should be contributing
to the gluon production in pA scattering is shown in \fig{zeropa}. The
diagram contains virtual interaction between the proton and the field
of the nucleus. Even though there is no interaction on the right hand
side of the diagram it is allowed by the ruled of the old fashioned
light cone perturbation theory, where the energy is not conserved in
the vertices \cite{meM,bl}. The diagram of \fig{zeropa} has a
remarkable feature in it, which was absent in the graphs shown in
\fig{zero}: merger of two produced gluons. In \fig{zeropa} two
Weizs\"{a}cker-Williams wave functions of the nucleus first interact
with the proton producing gluons, which then merge with each
other. This merging is different from the ones considered in
\fig{zero}. There one of the merging gluons was produced in the 
interaction with the proton while the other one was just given by the
non-interacting Weizs\"{a}cker-Williams wave function of the
nucleus. In \fig{zeropa} both merging gluons were first ``produced''
by the interactions and then merged together. Here again, since the
diagram of \fig{zeropa} does not contribute to the gluon production
cross section it has to either be zero or cancel with some other
diagrams. This gives us a very strong reason to conclude that the
gluons produced during the collision do not merge with each other at
later times in $A_+ = 0$ light cone gauge. Even a more general
conclusion can be conjectured: gluons produced in the interactions do
not interact with any other gluons afterwards in $A_+ = 0$ gauge.

To conclude we note that the physical interpretation of the scattering
process appears to be gauge dependent: in the covariant gauge case
considered in the previous section the gluon production was dominated
by multiple final state rescatterings. In the light cone gauge
multiple rescatterings vanish. The information about them is now
contained in the light cone wave function of the nucleus. The same
observation about the interplay of initial and final state
interactions was made in \cite{meM} for the case of current--nucleus
scattering with the current $j = - \frac{1}{4} F^{a \, 2}_{\mu\nu}$.

\section{Nucleus--Nucleus Collisions}

\begin{figure}
\begin{center}
\epsfxsize=15cm
\leavevmode
\hbox{ \epsffile{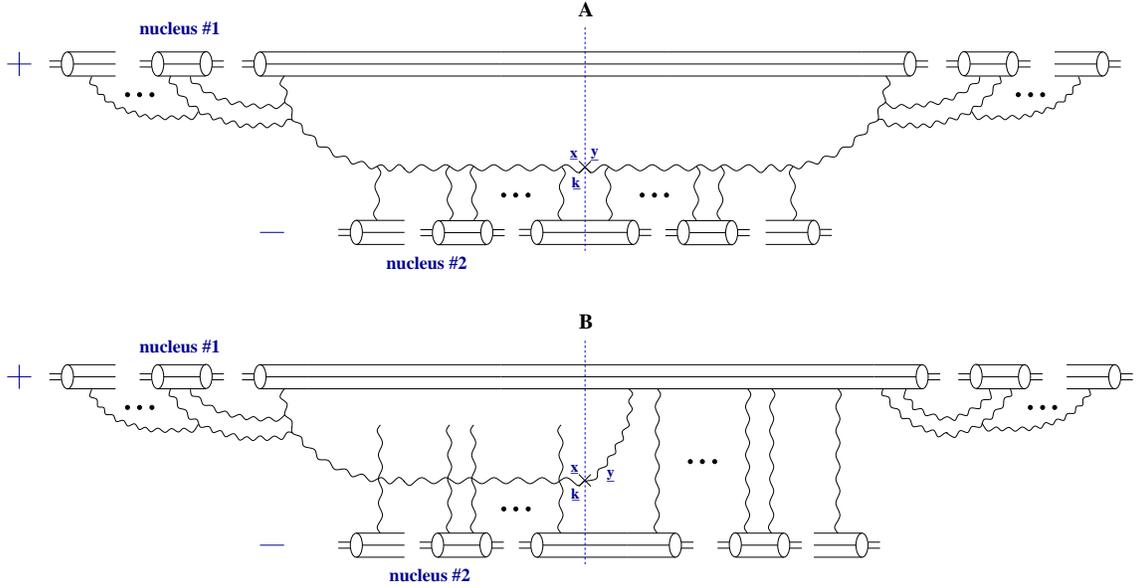}}
\end{center}
\caption{Diagrams contributing to the gluon production in nucleus-nucleus 
collisions in the $A_+ = 0$ light cone gauge.}
\label{aafig}
\end{figure}

Now we can employ the results we have obtained in Sect. III to write
down an ansatz for the distribution of produced gluons in
nucleus--nucleus collisions. Let us consider a head-on central
collision of two ultrarelativistic nuclei, as was shown in
\fig{coll}. We will be working in $A_+ = 0$ light cone gauge. The
calculations can be done either in the center of mass frame or in the
rest frame of one of the nuclei. We will work in the rest frame of the
second nucleus. The diagrams contributing to gluon production in AA
are depicted in \fig{aafig}. They are somewhat similar to the diagrams
of \fig{palc} and of \fig{pafig}. The incoming nucleus may or may not
have a Weizs\"{a}cker-Williams gluon in it. In the first case the
system multiply rescatters in the second nucleus at rest. This is
illustrated in \fig{aafig}A. Similar to pA case multiple rescatterings
between the first and the second nuclei cancel. Only the interactions
with the gluon survive.  The interference graph of the amplitude from
\fig{aafig}A and the amplitude where the gluon is emitted after the
interaction is shown in \fig{aafig}B. Analogous to pA case we work
with diagrams where the final state interactions are limited to
multiple rescatterings in the second nucleus and a single gluon
emission (or absorption) by the first nucleus. As was demonstrated in
Sect. III in the case of proton-nucleus scattering all other final
state interactions including ``produced'' gluons' merging cancel. Here
we argue that this also happens in the nucleus-nucleus collisions.

\begin{figure}
\begin{center}
\epsfxsize=8cm
\leavevmode
\hbox{ \epsffile{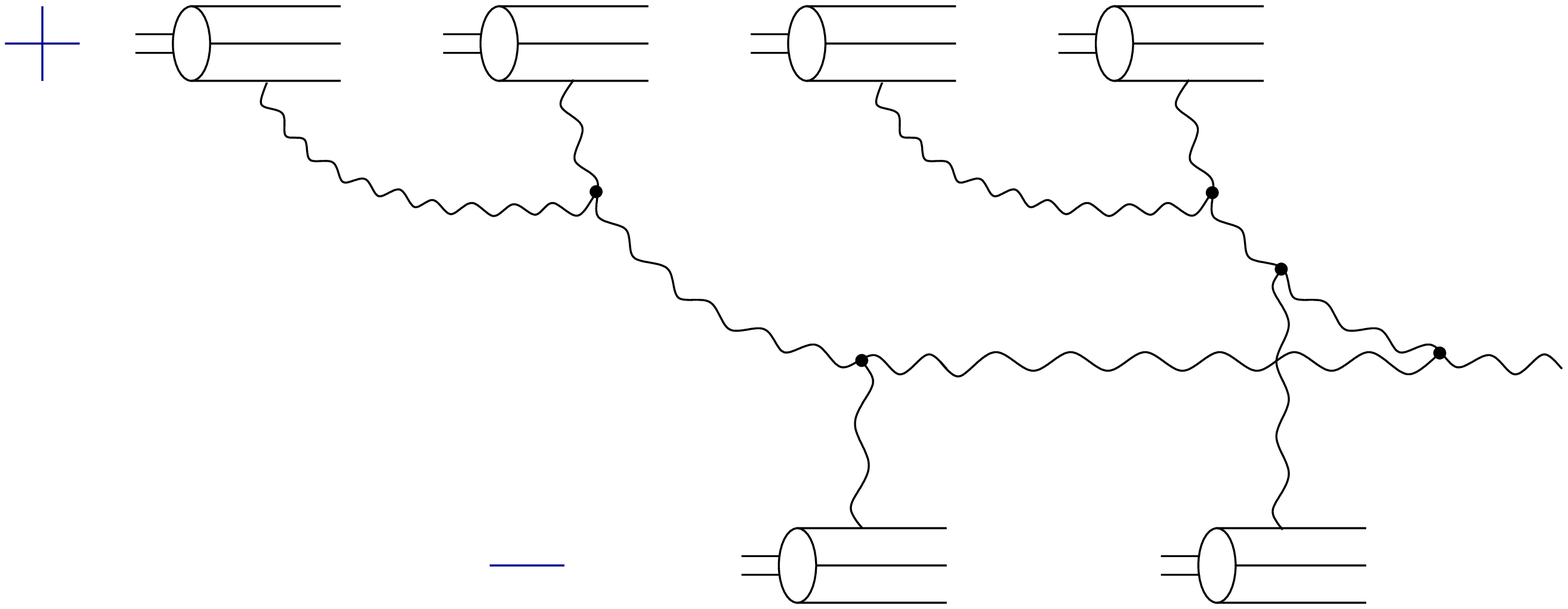}}
\end{center}
\caption{An example of a diagram which is not included in the gluon 
production mechanism of \fig{aafig}. To avoid possible confusion the
three gluon vertices in this diagram are marked by black dots.}
\label{zeroaa}
\end{figure}

In Sect. III, analyzing proton-nucleus scattering we concluded that
the diagrams where the gluon produced by interaction with the proton
merges with the non-Abelian Weizs\"{a}cker-Williams wave function of
the nucleus cancel. That allowed us to neglect similar diagrams in
\fig{aafig} above. However, there exists also a somewhat
different class of diagrams, one of which is shown in
\fig{zeroaa}. There the gluons that merge in the later stages of the
collision both were produced in the interaction of the
Weizs\"{a}cker-Williams wave function with the second nucleus. This
brings us back to the class of diagrams depicted in \fig{zeropa},
where there is also a merger of two ``produced'' gluons. Since the
diagrams of \fig{zeropa} canceled in the case of pA collisions we have
a strong reason to believe that the graphs of the type shown in
\fig{zeroaa} also cancel in AA collisions. Thus
\fig{aafig} contains all the diagrams that contribute to gluon production 
in nucleus-nucleus collisions.

We can calculate the diagrams in \fig{aafig} along the lines outlined
in the previous section. The contribution of the graphs of
\fig{aafig}A is obtained by a simple eikonalization of \eq{lc11}, 
and, correspondingly \fig{palc}A. The result for the number
distribution of the produced gluons yields
\be\label{aa11}
\frac{d N^{AA}_1}{d^2 k \, dy} \ = \ \int \frac{d^2 x \ 
d^2 y}{(2 \pi)^2} \, e^{i {\underline k} \cdot ({\underline x} -
{\underline y})}\, \frac{2}{\pi} \, \mbox{Tr} \, \left< {\underline
A}_1^{WW} ({\underline x}) \cdot {\underline A}_1^{WW} ({\underline
y}) \right>_1 \, \left(e^{ - ({\underline x} - {\underline y})^2
\, Q_{s2}^2 / 4 } - 1 \right).
\ee
Here the indices $1$ and $2$ denote the fields and the saturation
scales of the first and the second nuclei correspondingly, as well as
over which nucleus wave function the quantities are being
averaged. With the help of \eq{wwcorr} we rewrite \eq{aa11} as
\be\label{aa12}
\frac{d N^{AA}_1}{d^2 k \, dy} \ = \  - \frac{2 \, C_F}{\as \pi^2} \, \int \frac{d^2 x \ 
d^2 y}{(2 \pi)^2} \, e^{i {\underline k} \cdot ({\underline x} -
{\underline y})} \, \frac{1}{|{\underline x} - {\underline y}|^2} \,
\left(1 - e^{ - ({\underline x} - {\underline y})^2
\, Q_{s1}^2 / 4 } \right) \, \left(1 - e^{ - ({\underline x} - {\underline y})^2
\, Q_{s2}^2 / 4 } \right).
\ee
As was demonstrated in Appendix A of \cite{meM} multiple rescatterings
of \fig{aafig}B also just exponentiate. Therefore, in order to
calculate the contribution of the diagram of \fig{aafig}B we should
substitute
\be
\frac{\as \pi^2 N_c}{N_c^2 - 1} \, ({\underline x} - {\underline z})^2
\, xG (x, 1/({\underline x} - {\underline z})^2)
\ee
in \eq{lc21} by 
\be
1 - e^{ - ({\underline x} - {\underline z})^2 \, Q_{s2}^2 / 4 }.
\ee
Performing all the same integrations and averaging that were done in
obtaining \eq{lc22} we end up with
\be\label{aa21}
\frac{d N^{AA}_2}{d^2 k \, dy} \ = \  \frac{2 \, C_F}{\as \pi^2} \, \int \frac{d^2 x \ 
d^2 y}{(2 \pi)^3} \, e^{i {\underline k} \cdot ({\underline x} -
{\underline y})} \, d^2 b \, \frac{{\underline x}}{{\underline x}^2}
\cdot \frac{{\underline y}}{{\underline y}^2} \, \frac{1}{{\underline x}^2 
\, \ln \frac{1}{|{\underline x}| \mu}} \, \left(1 - e^{ - {\underline x}^2 
\, Q_{s1}^2 / 4 } \right) \, \left(1 - e^{ - {\underline x}^2
\, Q_{s2}^2 / 4 } \right).
\ee
To get the full answer we have to add the contribution of the diagram
which is complex conjugate to the one shown in \fig{aafig}B. Summing
up Eqs. (\ref{aa12}), (\ref{aa21}) and its complex conjugate, we
obtain the expression for the number distribution of gluons produced
in a heavy ion collision
\ben
\frac{d N^{AA}}{d^2 k \, d^2 b \, dy} \ = \ \frac{2 \, C_F}{\as \pi^2} \, 
\left\{  -  \int \frac{d^2 z}{(2 \pi)^2} \,  e^{i {\underline k} \cdot 
{\underline z}} \, \frac{1}{{\underline z}^2} \, \left(1 - e^{ -
{\underline z}^2 \, Q_{s1}^2 / 4 } \right) \, \left(1 - e^{ -
{\underline z}^2 \, Q_{s2}^2 / 4 } \right) + \right.
\een
\ben
+ \left. \int \frac{d^2 x \ d^2
y}{(2 \pi)^3} \, e^{i {\underline k} \cdot ({\underline x} -
{\underline y})} \,  \frac{{\underline x}}{{\underline x}^2}
\cdot \frac{{\underline y}}{{\underline y}^2} \,  \left[ \frac{1}{{\underline x}^2 
\, \ln \frac{1}{|{\underline x}| \mu}} \, \left(1 - e^{ - {\underline x}^2 
\, Q_{s1}^2 / 4 } \right) \, \left(1 - e^{ - {\underline x}^2
\, Q_{s2}^2 / 4 } \right) + \right. \right.
\een
\be\label{aasol}
+ \left. \left.  \frac{1}{{\underline y}^2 
\, \ln \frac{1}{|{\underline y}| \mu}} \, \left(1 - e^{ - {\underline y}^2 
\, Q_{s1}^2 / 4 } \right) \, \left(1 - e^{ - {\underline y}^2
\, Q_{s2}^2 / 4 } \right) \right] \right\}.
\ee
\eq{aasol} is our main result. It provides us with the number of gluons 
produced in a head-on zero impact parameter heavy ion collision per
unit transverse momentum phase space, per unit rapidity interval at
the given impact parameter $b$. \eq{aasol} includes the
nucleon-nucleon scattering result of \fig{lip} calculated in
\cite{kmw,med,mg} as well as the proton-nucleus scattering contribution 
of \fig{palc} found in \cite{meM}. We are going to explore the
properties of the distribution (\ref{aasol}) in the next section.

\section{Properties of the Classical Distribution}

Let us evaluate \eq{aasol} in the approximation in which we neglect
all the logarithms of the transverse coordinates \cite{meM}, since
logarithm is a slowly varying function and can be assumed to be a
constant compared to powers. That technically means putting $\ln
\frac{1}{|{\underline x}| \mu} \, \sim \, \ln \frac{1}{|{\underline y}| 
\mu} \, \sim \, 1 $ in \eq{aasol}. That also concerns terms like 
${\underline x}^2 \, Q_{s}^2$, which in general also have logarithms
of $|{\underline x}|$ in them, as follows from Eqs. (\ref{xqs}) and
(\ref{xg}). There we also put the logarithms to be of the order of
one, similar to how it was done in \cite{meM}. We have to note that
this approximation is good only for not very large transverse momenta
$k_\perp \lsim Q_s$. When gluon's momentum is large, $k_\perp \gg
Q_s$, the logarithms of the transverse coordinate are crucial for
deriving the correct asymptotics of the distribution function of
\eq{aasol}.

\begin{figure}
\begin{center}
\epsfxsize=9cm
\leavevmode
\hbox{ \epsffile{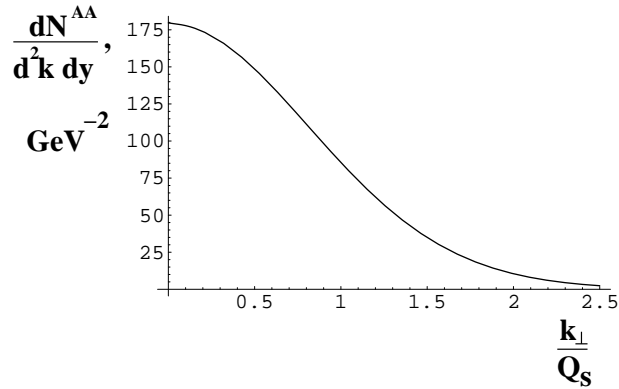}}
\end{center}
\caption{Distribution of the produced gluons given by \eq{dist2} for a 
central AA collision as a function of $k/Q_s$ with the transverse area
of the nuclei being $S_\perp = 50 \, \mbox{fm}^2$. The approximation
of \eq{aasol} is valid only for $k_\perp \, \ll \, Q_s$.}
\label{dist}
\end{figure}

Employing the fact that
\be
\int \frac{d^2 y}{(2 \pi)^2} \, e^{- i {\underline k} \cdot {\underline y}} 
\, \frac{{\underline y}}{{\underline y}^2} \,  = \, - \frac{i}{2 \pi} \, 
\frac{{\underline k}}{{\underline k}^2}
\ee
in the second term of \eq{aasol}, integrating over the angles of
${\underline z}$ and ${\underline x}$ and performing similar
integrations in the third term of \eq{aasol} we get
\be\label{dist1}
\frac{d N^{AA}}{d^2 k \, d^2 b \, dy} \ = \ \frac{C_F}{\as \pi^3} \, 
\int_0^\infty \frac{d x}{x} \, J_2 (k x) \, \left(1 - e^{ - {\underline x}^2 
\, Q_{s1}^2 / 4 } \right) \, \left(1 - e^{ - {\underline x}^2
\, Q_{s2}^2 / 4 } \right).
\ee
Integrating over $x$ in \eq{dist1} we find
\be\label{dist2}
\frac{d N^{AA}}{d^2 k \, d^2 b \, dy} \ = \ \frac{C_F}{\as 2 \pi^3} \, 
\frac{1}{{\underline k}^2} \, \left[ (Q_{s1}^2 + Q_{s2}^2) \, e^{- 
\frac{{\underline k}^2}{Q_{s1}^2 + Q_{s2}^2}} - Q_{s1}^2 \, e^{- 
\frac{{\underline k}^2}{Q_{s1}^2}} - Q_{s2}^2 \, e^{- 
\frac{{\underline k}^2}{Q_{s2}^2}}\right].
\ee
The distribution of \eq{dist2} is plotted in \fig{dist} as a function
of $k/Q_s$ for the case of two identical cylindrical nuclei with
$Q_{s1} = Q_{s2} = Q_s$ and with the cross sectional area $S_\perp \,
= \, \pi \, R^2 \, \approx \, 50 \, \mbox{fm}^2 \, \approx \, 1250 \,
\mbox{GeV}^{-2}$. Note again that \eq{dist2} is valid only in the not 
very large transverse momentum region $k_\perp \lsim Q_s$.

As one can see the distribution in \fig{dist} remains finite as
$k_\perp/Q_s \rightarrow 0$. If one takes the $k_\perp \, \ll \, Q_s$
limit of \eq{dist2} then the distribution goes to a constant
\be\label{ir}
\frac{d N^{AA}}{d^2 k \, d^2 b \, dy} \ \rightarrow \ \ \frac{C_F}{\as 2 \pi^3} 
\hspace*{1cm} \mbox{as} \hspace*{1cm} \frac{k_\perp}{Q_s} \, \ll \, 1.
\ee 
This means that the exact expression of \eq{aasol} may only have
logarithmic divergences in the infrared limit. This conclusion is very
interesting, since the initial conditions for heavy ion collisions
which are generated by pairwise interactions between nucleons in
nuclei without multiple rescatterings have a power-law divergences in
the infrared limit \cite{coll}. Even the pA gluon production cross
section of \eq{pasol} diverges as $1/k_\perp^2$ at small transverse
momenta because it includes multiple rescatterings in only one
nucleus, since one nucleus is involved in the scattering
process. Therefore finiteness of \eq{dist2} at small transverse
momenta demonstrates that multiple rescatterings are the reason the
hadronic and nuclear single particle inclusive production cross
sections remain finite in the soft momentum region.

Since the distribution of \eq{aasol} contains the lowest order in
$\as$ diagrams in it (see \fig{lip}) one readily derives that in the
$k_\perp/Q_s \rightarrow \infty$ limit the distribution falls off as
$1/k_\perp^4$ \cite{kmw,med,mg}
\be\label{uv}
\frac{d N^{AA}}{d^2 k \, d^2 b \, dy} \ \sim \ \frac{Q_{s1}^2 \, Q_{s2}^2}
{\as \, {\underline k}^4} \hspace*{1cm} \mbox{as} \hspace*{1cm}
\frac{k_\perp}{Q_s} \rightarrow \infty
\ee
which is a well-known perturbative result. As one can see
extrapolation of the usual perturbative expression of \eq{uv} into the
soft momentum region would lead to singularities and strong cutoff
dependence of the total number of the produced gluons. The multiple
rescatterings of \eq{aasol} resolve this problem.

A simple calculation shows that the typical transverse momentum of the
gluons in the distribution of \eq{dist2} is given by
\be\label{typk}
\left< {\underline k}^2 \right> \ = \ \frac{2 \, Q_{s1}^2 \, Q_{s2}^2 }{(Q_{s1}^2 
+ Q_{s2}^2) \, \ln (Q_{s1}^2 + Q_{s2}^2) - Q_{s1}^2 \, \ln Q_{s1}^2 -
Q_{s2}^2 \, \ln Q_{s2}^2 }.
\ee
For two identical cylindrical nuclei \eq{typk} gives
\be\label{typkn}
\left< {\underline k}^2 \right> \ = \ \frac{Q_{s}^2}{\ln 2}.
\ee
That is, the typical transverse momentum of the produced gluons is of
the order of the saturation scale, as was conjectured by Mueller in
\cite{Mueller2}. Since the saturation scale for a large nucleus scales
as $Q_s^2 \, \sim \, A^{1/3}$ with atomic number, as could be seen for
instance from \eq{qs}, it may get quite large, much larger than the
non-perturbative QCD scale $\Lambda_{QCD}$. Then most of the produced
gluons would have momenta high above $\Lambda_{QCD}$ which would
justify the use of perturbative QCD in the problem \cite{mv,Mueller2}.

Finally, the total number of gluons produced in the collision can be
found by integrating \eq{dist2} over $k_\perp$. The result yields 
\be
\frac{d N^{AA}}{d^2 b \, dy} \ = \ \frac{C_F}{\as 2 \pi^2} \, \left[ (Q_{s1}^2 
+ Q_{s2}^2) \, \ln (Q_{s1}^2 + Q_{s2}^2) - Q_{s1}^2 \, \ln Q_{s1}^2 -
Q_{s2}^2 \, \ln Q_{s2}^2 \right],
\ee
which, for the case of identical nuclei gives
\be\label{tot}
\frac{d N^{AA}}{d^2 b \, dy} \ = \ \frac{C_F \, Q_s^2 \, \ln 2}{\as \pi^2}.
\ee
In \cite{Mueller2} Mueller suggested that in a high energy nuclear
collision the gluons in the wave function of the incident nucleus get
liberated by the interactions with the nucleus at rest. The coherence
of the incoming nucleus gluonic wave function is broken by the second
nucleus. Thus the total number of produced gluons should be
proportional to the total number of gluons in the wave function of one
of the nuclei before the collision with the proportionality
coefficient $c$, which should be of order one
\cite{Mueller2}. Therefore one may write
\cite{Mueller2}
\be
\frac{d N^{AA}}{d^2 b \, dy} \ = \ c \ \frac{d N^{WW}}{d^2 b \, dy} \,
= \, c \ \frac{2}{\pi} \, \mbox{Tr} \left< {\underline A}^{WW}
({\underline x}) \cdot {\underline A}^{WW} ({\underline x}) \right>,
\ee
which, using \eq{wwcorr} can be rewritten as
\be\label{totmu}
\frac{d N^{AA}}{d^2 b \, dy} \ = \ c \ \frac{C_F \, Q_s^2}{\as \, 2 \, \pi^2}.
\ee
Comparing \eq{totmu} to \eq{tot} we conclude that
\be\label{coef}
c \ = \ 2 \, \ln 2 \ \approx \ 1.39 ,
\ee
which is very close to the result of the numerical estimates of
Krasnitz and Venugopalan giving $c \, = \, 1.29 \, \pm \, 0.09$
\cite{kv}. The obtained value for the ``gluon liberation'' coefficient 
$c = 2 \ln 2$ is close to one, as was originally suggested by Mueller
\cite{Mueller2}.

\section{Discussion}

\eq{tot} allows us to estimate the saturation scale $Q_s$ knowing the 
multiplicity of produced particles. Of course one should be careful in
interpreting this estimate, since at high energies the purely
classical picture considered here breaks down and quantum corrections
bringing in powers of $\as \ln s$ become important. It has been
conjectured though \cite{mv,klm} that these corrections would not
change \eq{tot} and would only (considerably) increase the value of
the saturation scale $Q_s$ on the right hand side of it. Another issue
one should be worried about in this kind of an estimate is that the
classical gluon production picture presented here does not include the
interactions at late times, which may lead to thermalization of quarks
and gluons produced, and may also modify the total number of gluons
\cite{bmss}. \eq{tot} gives us the total number of gluons immediately 
after the collision, which may be different from what the detectors
count at the end due to the importance of $2 \rightarrow 3$ and $3
\rightarrow 2$ processes at the later stages of the collision
\cite{bmss}. Also \eq{tot} gives us the total number of gluons
produced, which is not quite equal to the total number of pions, kaons
and other hadrons observed in the detector. Here we will just assume
that due to entropy conservation the numbers are very close to each
other \cite{klm}.  Keeping all the above mentioned restrictions in
mind we may nevertheless try to estimate $Q_s$ in our classical
picture here using the newly emerging RHIC data \cite{pho}. PHOBOS
experiment has measured total charge multiplicity per unit
pseudorapidity in Au$+$Au collisions yielding the result \cite{pho}
\be\label{pmult}
\frac{dN^{Au+Au}_{ch}}{d \eta} \ = \ 555 \, \pm \, 12 (\mbox{stat}) \, \pm \, 
35 (\mbox{syst})
\ee
at the center of mass energy $\sqrt{s} \, = \, 130 \, \mbox{AGeV}$. In
our crude estimate we will multiply the number given in \eq{pmult} by
$3/2$ to account for charge neutral particles and use the resulting
number as a lower bound estimate of $dN/dy$ in \eq{tot}. ($dN/dy$ is a
little larger than $dN/d\eta$.) Again we use a cylindrical nucleus
approximation with the cross sectional area $S_\perp \, = \, 50 \,
\mbox{fm}^2$. We assume that the strong coupling constant is $\as \, 
\approx \, 0.3$. (In general $\as \, = \, \as (Q_s^2)$ and we have 
to treat \eq{tot} as an implicit equation.) The result for the
saturation scale is
\be\label{qsn}
Q_s^2 \ \approx \ 2.1 \, \mbox{GeV}^2 \hspace*{1cm} \mbox{for} \ Au+Au
\ \mbox{at} \ \sqrt{s} \, = \, 130 \, \mbox{AGeV},
\ee
which is close to and even a little larger than the estimate of
\cite{ekrt}.  The saturation scale of \eq{qsn} appears to be
marginally in the perturbative region. As energy of the RHIC beam
reaches $200 \, \mbox{AGeV}$ the particle multiplicity will increase
too, leading to an even larger saturation scale, which would make the
use of perturbative QCD at RHIC even better justified.

To summarize the results of this paper we repeat again that we have
derived the classical distribution of gluons produced in the
ultrarelativistic heavy ion collision (\eq{aasol}), thus constructing
classical initial conditions for the evolution of the gluon system
leading to a possible gluon thermalization. It would be very
interesting and important to analyze the subsequent evolution of the
gluonic system in the framework of McLerran-Venugopalan model and see
whether the onset of thermalization is possible before the system
falls apart and to what experimental consequences that would
lead. Important first steps in that direction have already been made
\cite{dg,bmss,Mueller2}. \eq{aasol} can also be applied to describe 
minijet production in the proton-proton collisions at very high
energies \cite{kmw,mg}. The proton's high energy wave function
consists of many sea partons which may serve as color charge sources
for the classical field similar to nucleons in the nuclear case
\cite{mv,JKLW,bal}. We have derived a simplified expression for the distribution of
\eq{aasol} which is given in \eq{dist2}. We have demonstrated that the
distribution is finite in the soft transverse momentum region
(\eq{ir}) and approaches the usual perturbative result when the
transverse momentum becomes large (\eq{uv}). We have shown that the
typical transverse momentum of the produced gluons in the collision of
two identical nuclei is of the order of the saturation scale $Q_s$
(\eq{typkn}). Finally we have observed that the total number of the
produced gluons is proportional to the total number of the gluons in
the nuclear wave function with the proportionality coefficient $c = 2
\ln 2$ (\eq{coef}).

\section*{Acknowledgements}

I would like to thank Ian Balitsky, Larry McLerran, Jerry Miller, Al
Mueller and Raju Venugopalan for many informative discussions on the
subject. This work has been supported in part by the U.S. Department
of Energy under Grant No. DE-FG03-97ER41014.

\end{document}